\documentclass[12pt,aps,prl,floats,graphicx,euscript,amsmath,amssymb,
onecolumn]{revtex4}
\begin{document}

\title{Fermi point in graphene as a monopole in momentum space}
\author{M.A.Zubkov\\
ITEP, B.Cheremushkinskaya 25, Moscow, 117259, Russia }

 \rightline{\small ITEP-LAT/2011-13 }

\newcommand{\br}{{\bf r}}
\newcommand{\bu}{{\bf \delta}}
\newcommand{\bk}{{\bf k}}
\newcommand{\bq}{{\bf q}}
\def\({\left(}
\def\){\right)}
\def\[{\left[}
\def\]{\right]}


\begin{abstract}
{ We consider the effective field theory of graphene monolayer with the Coulomb
interaction between fermions  taken into account. The gauge field in momentum
space is introduced. The position of the Fermi point coincides with the
position of the corresponding monopole. The procedure of extracting such
monopoles during lattice simulations is suggested.}
\end{abstract}

\maketitle

\section{Introduction}

Graphene is a unique $2+1 $ - dimensional nonrelativistic system that shares
common properties with relativistic quantum field theory. In particular, in the
effective field theory of graphene the massless Dirac spinors appear
\cite{CastroNeto:2009zz,Fialkovsky,Semenoff:1984dq,Wallace}. When the Coulomb
interaction is taken into account, the effective Lorentz symmetry is broken.
The phase structure of the model may be changed when the external conditions
are changed (that may lead, say, to the change of Fermi velocity $v_F$)
\cite{Araki,rev_sym,Semenoff,Son,Timo}. Change of the phase structure of the
model must be accompanied with the deformation of the momentum space topology
\cite{VolovikBook,VolovikLect}. Therefore, it is important to investigate
various topological invariants in momentum space of the effective field model.

In general in $3D$ the Fermi points are not topologically stable
\cite{VolovikBook}. This is because $\pi_2(GL(N,C)) = 0$ for $N\ge 2$. The
$N\times N$ Green function in momentum space belongs to $GL(N,C)$. That's why
topological triviality of mapping $S_2 \rightarrow GL(N,C)$ does not allow
topological stability of the Green function's poles in general case. However,
if a certain symmetry is present that reduces the size of the space of the
Green functions, the topological stability becomes possible
\cite{VolovikLect,rev_sym}. In particular, the $3D$ model of graphene monolayer
has such a symmetry that effectively reduces space of the Green functions
considerably. As a result, the topological invariant ${\cal N}_2$ appears. This
invariant is expressed through Green function at zero frequency $\omega = 0$
and is an integral over the closed contour $\cal C$ around Fermi point in the
plane of $2D$ momenta $\bf p$ (see, for example, \cite{VolovikLect}). The
important advantage of the existence of this invariant is that the pole of the
Green function cannot disappear without a phase transition. It is worth
mentioning, however, that this construction is less natural than that of the
invariant ${\cal N}_3$ of $4D$ theory \cite{VolovikBook}. This is because only
the structure of $\omega = 0$ plane in momentum space is reflected by ${\cal
N}_2$. Moreover, the given construction was introduced when Coloumb interaction
between quasiparticles is neglected, and in the case, when these interactions
are present, it requires an additional investigation.

In this paper we extend the construction of ${\cal N}_2$ to the effective field
model of graphene in such a way that the topological invariant is written as an
integral over the surface in $3D$ momentum space $(\omega, {\bf p})$. In the
form presented here this invariant works also for the case, when the Coulomb
interaction is present. We show that when the Green function is smooth enough,
our construction can be reduced to the original construction of ${\cal N}_2$.
In addition, we present the definition of the gauge field in momentum space
such that the positions of the corresponding monopoles coincide with the
positions of the poles or zeros of the Green function. This construction is
intended mainly to be used during lattice simulations. We also suggest the
procedure of extraction the monopoles in momentum space for the lattice
discretization with staggered fermions.

\section{The field theoretical effective model for graphene}
\label{IntroductionSec}


The low energy effective model of graphene may be derived
\cite{Semenoff:1984dq,Wallace,CastroNeto:2009zz} starting from the simple non -
relativistic Hamiltonian that describes the interactions of electrons that
belong to neighbor Carbon atoms. The carbon atoms of graphene form a honeycomb
lattice with two sublattices A and B (or the triangular form). Further we
denote the lattice spacing by $a$. Let us introduce vectors that connect a
vertex of the sublattice A to its neighbors (that belong to the sublattice B):
$ {\bf l}_1= (-a,0),\qquad {\bf l}_2 = (a/2,a\sqrt{3}/2),\qquad {\bf l}_3=
(a/2,-a\sqrt{3}/2)$. The Hamiltonian has the form
\begin{equation}
H=-t\sum_{\alpha\in A}\sum_{j=1}^3
    \Bigl(\psi^\dag (\br_\alpha) \psi(\br_\alpha + {\bf l}_j)
        + \psi^\dag (\br_\alpha +{\bf l}_j)\psi(\br_\alpha)\Bigr)\,,\label{H1}
\end{equation}
Here $t$ is the hopping parameter,  operator $\psi^\dag$ creates electrons at
the points of the lattice.

Let us define  two electron fields  in momentum space that correspond to two
sublattices:
\begin{equation}
\psi_A({\bf k})  =  \frac{1}{V} \sum_{\alpha \in A} \psi(r_{\alpha}) e^{-i {\bf
k} (r_{\alpha}+{\bf l}_1)}, \quad  \psi_B({\bf k})   = \frac{1}{V} \sum_{\beta
\in B} \psi(r_{\beta}) e^{-i {\bf k} r_{\beta}}
\end{equation}
Here $V$ is the number of points in the sublattice $A$. The Brillouin zone is a
hexagon with opposite sides identified. There are two different vertices of the
hexagon that are denoted $K_+$, $K_-$. Quasiparticle energy vanishes at these
points. We expand $\psi$ around $K_+$, $K_-$,  denote $\psi^{\pm}_{A,B}(\bq
)\equiv \psi_{A,B}(K_\pm +\bq)$ and introduce the 4 - component field $ \psi =
\left( \psi_A^+, \psi_B^+,  \psi_A^-,  \psi_A^- \right)^T $.

At low energy the effective field theory appears. Taking the Fourier transform
from ${\bf q}$ to the coordinate space we come to the field - theoretic
formulation of the model:
\begin{equation}
H =  \int d^2 x \psi^\dag({\bf x}) \hat{D} \psi_A({\bf x}),\label{D}
\end{equation}
where $\hat{D}$ has the form of the usual Dirac operator taken on the $2D$
hypersurface $x_3 = 0$: $\hat{D} = - i v_F\gamma^0 \gamma^a
\partial_a$, $a=1,2$, where $v_F =(3ta)/2$ is the Fermi velocity (that is about $1/300$).
Here   $\gamma$ are the gamma - matrices in the representation to be specified
below. Let us remind that we started from the nonrelativistic Hamiltonian and
completely disregarded spin degrees of freedom. Now we take them into account
adding a new index to the field $\psi$. We assume it has two spin components.
In hamiltonian (\ref{D}) gamma - matrices act on the pseudospin index while the
true spin operator does not enter the Hamiltonian.

We consider the interaction between quasiparticles due to the photon exchange
($A$ is the $3 + 1$ electromagnetic field). Let us perform the Wick rotation,
the rescaling of time, and gauge fields: $ t \rightarrow  i x^4/v_F \quad A^0
\rightarrow i \sqrt{v_F} A^4 \quad \bar{A} \rightarrow
 \frac{1}{\sqrt{v_F}} \bar{A}$.

Further we denote $g = e/\sqrt{v_F}$. Therefore, the analogue of the fine
structure constant is $\alpha_F = \alpha/v_F \sim 300/137 \sim 2$. We also
introduce Euclidean Dirac matrices that satisfy $\{\Gamma_i,\Gamma_j\} = 2
\delta_{ij}$:
\begin{eqnarray}
&& \Gamma^4 = \gamma^0=\left( \begin{array}{cc} \sigma_3 & 0 \\ 0 & \sigma_3
\end{array} \right),\qquad \Gamma^1 = i\gamma^1=\left( \begin{array}{cc} -\sigma_1& 0 \\ 0
& -\sigma_1
\end{array} \right),\quad \Gamma^2 = i\gamma^2=\left( \begin{array}{cc} -\sigma_2& 0 \\ 0
& \sigma_2 \end{array} \right) \nonumber\\ && \Gamma^3 = \left(
\begin{array}{cc} 0 &-\sigma_2  \\  -\sigma_2 & 0 \end{array} \right), \quad
\Gamma^5 = \left( \begin{array}{cc} 0 & -i\sigma_2 \\  i\sigma_2 & 0
\end{array} \right),  \quad i\Gamma^3 \Gamma^5 =\left(
\begin{array}{cc} 1 & 0  \\  0 &- 1  \end{array} \right), \quad \bar{\psi} =
\psi^+ \Gamma^4 \label{Gamma}
\end{eqnarray}

 We introduce finite temperature $T$ via taking an integral over
$x^4$ within the interval $[0, \frac{v_F}{T}]$ and adopting the periodic in
$x^4$ boundary conditions. The chemical potential is assumed to be equal to
zero. Due to $v_F \sim 1/300 << 1$ the fluctuations of $A_a$ are suppressed and
we neglect them in the functional integral. We arrive at the partition
function:
\begin{eqnarray}
&& Z   = \int D\bar{\psi}D\psi D A {\rm exp} \Bigl( - \frac{1}{2}\int d^4x
[\partial_{I} A_{4}]^2 - \int d^3x \bar{\psi}^A([\partial_4 - i g A_4 ]
\Gamma^4 +
\partial_a  \Gamma^a)\psi_A \Bigr),\nonumber\\
&& a = 1,2; \, I, J = 1,2,3
\end{eqnarray}
Here index $A = 1,2$ belongs to the spin degrees of freedom.

\section{Green functions}

It is worth mentioning that the Green function has to be considered in a
certain gauge. The gauge freedom of the system corresponds to the
transformation $ A_4 \rightarrow A_4 + \partial_4 \alpha(x^4) \quad \psi
\rightarrow e^{i\alpha} \psi $. We may fix this gauge freedom via implying a
certain gauge fixing condition. For example we may choose the condition
$A_4(x^4, {\bf z}) = 0$ for a certain $3D$ position $\bf z$. The model at
finite temperature, i.e. with periodic boundary conditions along $x^4$ should
be considered with care. When the system is considered in lattice
regularization, the value of $A_4$ on a certain point $(x^4_0, {\bf z})$ must
not be fixed. This choice of the gauge might appear to break general properties
of the Green functions (see Appendix) as it introduces the selected point in
plane $x - y$. We may instead fix another gauge minimizing the functional $\int
A_4^2 d^4 x$ with respect to the gauge transformations. Further we imply that
the given gauge is fixed and the gauge fixing condition is inserted into the
functional measure over $A_4$.

 Fermion Green function has the form:
\begin{eqnarray}
&& {\bf G}  =  \langle {\psi}^{\dag}_x \psi_{y} \rangle   = \frac{1}{Z}\int
D{\psi}^{\dag}D\psi D A {\psi}^{\dag}_x \psi_{y}{\rm exp} \Bigl( -
\frac{1}{2}\int d^4x [\partial_{I} A_{4}]^2 \nonumber\\&& - \int d^3x
\bar{\psi}([\partial_4 - i g A_4] \Gamma^4 +
\partial_a  \Gamma^a\psi \Bigr),\nonumber\\
&& a = 1,2; \, I, J = 1,2,3 \label{Green}
\end{eqnarray}

In order to reveal the $3D$ nature of the system let us consider the following
representation of the spinor field:
\begin{equation}
\psi\equiv \left(\begin{array}{c} \chi_+\\  \sigma_2\chi_- \end{array}\right)
\end{equation}

In terms of $\chi_+$ and $\chi_-$ the Green functions are:
\begin{eqnarray}
&& {\cal G}_{\pm \pm}  =   \frac{1}{Z}\int D{\chi}^{\dag}D\chi D A
{\chi}^{\dag}_{\pm}(x) \chi_{\pm}(y){\rm exp} \Bigl( - \frac{1}{2}\int d^4x
[\partial_{I} A_{4}]^2 \nonumber\\&& - \int d^3x
{\chi_-}^{\dag}\sigma_3([\partial_4 - i g A_4]\sigma_3 - \partial_1  \sigma_1 -
\partial_2 \sigma_2)\chi_-\Bigr)\nonumber\\&& - \int d^3x
{\chi_+}^{\dag}\sigma_3([\partial_4 - i g A_4]\sigma_3 -  \partial_1  \sigma_1
-   \partial_2  \sigma_2)\chi_+\Bigr)
\end{eqnarray}
We have, obviously, ${\cal G}_{+ +} = {\cal G}_{- -} = {\cal G}$. At the same
time ${\cal G}_{\pm \mp} = \langle {\chi}^{\dag}_{\pm}(x) \chi_{\mp}(y) \rangle
= 0 $. We also imply that the Green function is diagonal in spin index. That's
why $\cal G$ can be understood as the $2\times 2$ matrix. On the language of
$\chi_{\pm}$ different $\Gamma_3$ chiralities correspond to the states with
$\chi_{+}=\pm \chi_{-}$. Different $\Gamma_5$ chiralities correspond to the
states with $\chi_{+}=\pm i \chi_{i}$. $i\Gamma_3\Gamma_5$ chiralities
correspond to $\chi_{\pm} = 0$. In momentum space the $2\times 2$ matrix $\cal
G$ can be expressed as
\begin{equation}
{\cal G}(\omega, {\bf p}) = \int d^3x {\cal G}(0,x)e^{i \omega x^4 + i ({\bf p}
{\bf x})} = i \{g_0(\omega, {\bf p}) + { g}_a(\omega, {\bf p})
\sigma^a\}\sigma_3, \, a = 1,2,3 \label{Green0}
\end{equation}
Here vectors ${\bf p}, {\bf x}$ are two - component.

Direct calculation gives
\begin{eqnarray}
 {\cal G}&  = &  \frac{1}{Z} \int  D A  {\rm exp} \Bigl( -
\frac{1}{2}\int d^4x [\partial_{I} A_{4}]^2 \Bigr) \nonumber\\&& {\rm Det}^2
\Bigl(i[\partial_4 - i g A_4]\sigma_3 - i\partial_1  \sigma_1 - i\partial_2
\sigma_2\Bigr) \nonumber\\&& \frac{i}{i[\partial_4 - i g A_4]\sigma_3 - i
\partial_1  \sigma_1 -i   \partial_2
\sigma_2} \sigma_3
\end{eqnarray}

Operator $Q = i[\partial_4 - i g A_4]\sigma_3 - i[\partial_1 ] \sigma_1 - i
[\partial_2] \sigma_2$ is Hermitian for any real $A_4$. That's why we come to
the conclusion that the operator $i {\cal G} \sigma_3$ is also Hermitian. This
means that the functions $g_a(\omega, {\bf p}), a = 0,1,2,3$ are real. As a
results $-i {\cal G} \sigma_3$ belongs to $u(2)$.  Considering symmetries of
the Green function, we come to the following form of $\cal G$ (see Appendix):
\begin{eqnarray}
g_0(\omega, {\bf p})& = & 0, \quad g_3(\omega, {\bf p}) =
\tilde{f}\Bigl(\omega, |{\bf p}|^2\Bigr), \nonumber\\ {\bf g}(\omega, {\bf p})
& = & {\bf p} \tilde{h}\Bigl(\omega^2,|{\bf p}|^2 \Bigr), \,
\tilde{f}\Bigl(\omega, |{\bf p}|^2\Bigr) = -\tilde{f}\Bigl(-\omega, |{\bf
p}|^2\Bigr)
\end{eqnarray}
That's why $i {\cal G} \sigma_3 \in su(2)$. If in addition, the scale
invariance is not broken (in particular, $T = 0$), and the functions
$\tilde{f}, \tilde{h}$ are smooth enough, we have the further simplification:
\begin{equation}
g_3(\omega, {\bf p}) = \frac{\omega}{\omega^2+|{\bf p}|^2}
f\Bigl(\frac{\omega^2}{|{\bf p}|^2}\Bigr), \quad {\bf g}(\omega, {\bf p}) =
\frac{\bf p}{\omega^2+|{\bf p}|^2} \, h\Bigl(\frac{\omega^2}{|{\bf p}|^2}\Bigr)
\end{equation}

\section{Topological invariant at $\omega = 0$}

In some publications (see, for example, \cite{VolovikLect}) the following
expression has been considered that is shown to be a topological invariant both
with and without external magnetic field when the interaction with $A_4$ is
switched off:
\begin{equation}
{\cal N}_2 = \frac{1}{4 \pi i} {\rm Tr}\, \sigma_3 \int_{\cal C}  {\cal G} d
{\cal G}^{-1}\label{N20}
\end{equation}
Here contour $\cal C$ around the Fermi point (the pole of $\cal G$) is taken in
the $\omega = 0$ plane. For the noninteracting fermions we have ${\cal N}_2 =
1$. Further, if the interactions are introduced, the Green function is changed:
${\cal G} \rightarrow {\cal G} + \delta {\cal G}$. If the interactions are such
that $\{\delta {\cal G}(0,{\bf p}), \sigma_3\} = 0$, then $\delta {\cal N}_2 =
0$. This means that ${\cal N}_2 = 1$ until the phase transition is encountered.
For example, if the external magnetic field in $z$ - direction is introduced we
 have $\{\delta {\cal G}(0,{\bf p}), \sigma_3\} = 0$ (see, for example,
 \cite{Hatsugai}). At a first look, it is not obvious, that with the Coulomb interaction
 turned on ${\cal N}_2$ remains the topological invariant.

However, using the above mentioned symmetry considerations we rewrite this
function in the absence of external fields as follows:

\begin{equation}
{\cal N}_2  = -\frac{1}{4 \pi i} {\rm Tr}\, \sigma_3 \int_{\cal C}  \frac{ (d
g_3 \sigma_3 + (d {\bf g}, \sigma ))(g_3 \sigma_3 + ({\bf g}, \sigma ) )
}{g_3^2 + {\bf g}^2} = \frac{1}{2 \pi } \int_{\cal C} \frac{\epsilon_{ab} n^a d
n^b}{1+\frac{g_3^2}{{\bf g}^2}}
\end{equation}
where $n^1 = p^1/\sqrt{[p^1]^2 + [p^2]^2},\, n^2 = p^2/\sqrt{[p^1]^2 +
[p^2]^2}$, and it is implied that $\tilde{h}(0,{\bf p}^2)\ne 0$. As it was
mentioned above, when the functions $g_a$ are smooth enough, we have $g_3(0,
{\bf p}) = 0$ and, therefore, again ${\cal N}_2 = 1$. This means that the pole
of the Green function (the Fermi point) is topologically stable if the
symmetries considered above take place.

\section{Topological invariant in space $(\omega,{\bf p})$}

Below we generalize the construction of the topological invariant ${\cal N}_2$
considered above. The resulting construction uses the Green function defined on
the surface that encloses the Fermi point in $\omega - {\bf p}$ space. The
considered construction also works for nonzero $g_3(0,{\bf p})$.  Let us define
the function in momentum space
\begin{equation}
{\cal H} = \frac{{\cal G} \sigma_3 }{\sqrt{\frac{1}{2}{\rm Tr}\, ({\cal
G}\sigma_3)^2}}
\end{equation}

We can express ${\cal H}$ through the functions $g_a$ mentioned above: $ {\cal
H} = n_a \sigma_a,\quad n_a = g_a/|g|, \quad |g| = \sqrt{g_a g_a}, \quad a =
1,2,3$. Now let us consider the following integral over closed surface $\Sigma$
in momentum space such that $\cal G$ does not have poles on $\Sigma$:

\begin{equation}
{\cal N}_2 = \frac{1}{16 \pi i} {\rm Tr}\, \int_{\Sigma} {\cal H}\, d {\cal H}
\wedge d {\cal H} = \frac{1}{8 \pi } \epsilon_{abc}\, \int_{\Sigma} n^a\, d n^b
\wedge d n^c \label{N2}
\end{equation}

The given expression (\ref{N2}) for the invariant ${\cal N}_2$ is, obviously,
reduced to (\ref{N20}) in the case, when $g_3(0, {\bf p}) = 0$. Without
interactions and without external magnetic field $\cal G$ has the pole at ${\bf
p} = \omega = 0$ that corresponds to the Fermi point. In this case $n_3(\omega,
{\bf p}) = \frac{\omega}{\sqrt{\omega^2+{\bf p}^2}},\,  {\bf n}(\omega, {\bf
p}) = \frac{\bf p}{\sqrt{\omega^2+{\bf p}^2}} $, and ${\cal N}_2 = 1$ for any
surface that encloses the pole. When the interaction with the electromagnetic
field is turned on, the value of ${\cal N}_2$ for the surface that encloses
this pole remains equal to unity until a phase transition is encountered. The
important advantage of the given formulation is that we already do not need the
condition $g_3(0, {\bf p}) = 0$ to be satisfied. We only need $g_0 = 0$. The
situation, when $g_0 = 0$ and $g_3(0, {\bf p}) \ne 0$ may appear in the other
$2+1$ systems or even in the effective field model of graphene for the
inhomogenious gauge or when some of the symmetries are broken dynamically.

\section{Fermi point as a monopole}

As it was explained above, $\tilde{\cal G} = - i {\cal G} \sigma_3 \in su(2)$
out of the region, where $\cal G$ has  poles. We can  diagonalize $\tilde{\cal
G}$ via $SU(2)/U(1)$ transformations:
\begin{equation}
\tilde{\cal G} = V^{\dag}(\sqrt{g_3^2 + {\bf g}^2} \sigma_3)V
\end{equation}
$V$ is defined up to the $U(1)$ transformation $V \rightarrow e^{\alpha
\sigma_3} V$. That's why here $V \in SU(2)/U(1)\sim S_2$. We can choose $V$ to
be smooth on the surface $\Sigma$ except for a small vicinity $\Omega$ of a
certain point. We have $\pi_2(SU(2)/U(1)) = Z$. Actually the invariant ${\cal
N}_2$ is equal to the degree of the mapping $S_2 \rightarrow SU(2)/U(1)$:
\begin{eqnarray}
{\cal N}_2 & =& \frac{1}{16 \pi i } {\rm Tr}\, \int_{\Sigma - \Omega} V^{\dag}
\sigma_3 V
d [V^{\dag} \sigma_3 V] \wedge d [V^{\dag} \sigma_3 V]\nonumber\\
&=& - \frac{1}{4 \pi i} {\rm Tr}\, \int_{\Sigma-\Omega}  \sigma_3 d V \wedge d
V^{\dag} = - \frac{1}{4 \pi i} {\rm Tr}\, \int_{\Sigma-\Omega}  \sigma_3
d[ V d V^{\dag}]\nonumber\\
& = &  \frac{1}{4 \pi i} {\rm Tr}\, \int_{\partial \Omega}  \sigma_3  V d
V^{\dag}
\end{eqnarray}
Now we define the gauge field in momentum space ${\cal B} = -i V d V^{\dag}$.
$\cal B$ is smooth everywhere except for the string ended at the position of
the pole of $\cal G$. The field strength of $\cal B$ vanishes everywhere except
for the mentioned string.  The position of the string (but not the positions of
its ends) can be changed by the $U(1)$ transformations $V \rightarrow e^{\alpha
\sigma_3} V$. The third component of the gauge field $B = \frac{1}{2} {\rm
Tr}\, {\cal B} \sigma_3$ is the $U(1)$ field. The position of the corresponding
Dirac monopole coincides with the pole (or zero) of $\cal G$. The position of
the Fermi point without interactions coincides with the position of the
monopole constructed of $B$ in momentum space. The position of the antimonopole
coincides with the zero of $\cal G$ (placed at the infinity). Monopole and
antimonopole are connected by the Dirac string. This pattern cannot disappear
until the phase transition is encountered.

\section{${\cal N}_2$ in $4D$ notations}

In $4D$ notations  Green function (\ref{Green}) has the form:
\begin{equation}
{\bf G} = \left( \begin{array}{cc}{\cal G} & 0 \\ 0 & \sigma_2 {\cal G}
\sigma_2
\end{array}\right) = i \left( \begin{array}{cc}\tilde{\cal G} & 0 \\ 0 & -\sigma_2 \tilde{\cal G}
\sigma_2
\end{array}\right)\Gamma_4\label{G4D}
\end{equation}
Again, we define the function in momentum space: ${\bf H} = \frac{-i{\bf
G}\,\Gamma_4 }{\sqrt{\frac{1}{4}{-\rm Tr}\, ({\bf G}\,\Gamma_4)^2 }}$. We can
express ${\bf H}$ through three real functions $g_a$ mentioned above: $ {\bf H}
= - n_1 \Gamma_1 - n_2 \Gamma_2 + n_3 \Gamma_4 ,\quad n_a = g_a/|g|, \quad |g|
= \sqrt{g_a g_a}, \quad a = 1,2,3$. Now the topological invariant can be
expressed as
\begin{equation}
{\cal N}_2 = \frac{1}{32 \pi } {\rm Tr}\, \int_{\Sigma} {\bf H}\, d {\bf H}
\wedge d {\bf H} \, \Gamma_3 \Gamma_5 = \frac{1}{8 \pi } \epsilon_{abc}\,
\int_{\Sigma} n^a\, d n^b \wedge d n^c
\end{equation}

We denote $\tilde{\bf G} = - i {\bf G} \Gamma_4$. $\tilde{\bf G}$ can be
diagonalized via the $SO(4)/(SU(2)\otimes U(1))$ transformations:
\begin{equation}
\tilde{\bf G} = {\bf V}^{\dag}( \sqrt{g_3^2 + {\bf g}^2} \Gamma_4){\bf V}
\end{equation}
Here $ {\bf V} = \left( \begin{array}{cc}{V} & 0 \\ 0 & \sigma_2 V \sigma_2
\end{array}\right)$.
$V$ can be chosen in the form:
\begin{equation}
V = {\rm exp}\Bigl( \frac{i(n_2 \sigma_1 - n_1 \sigma_2){\rm arccos} \, n_3
}{2\sqrt{1-n_3^2}}\Bigr)\label{V}
\end{equation}

\section{Momentum space topology of lattice regularized model}

Staggered fermions are unique for the graphene monolayer because in this
regularization the doublers of the one - component fermion play the role of the
components of two Dirac spinors. This regularization has been used in practical
numerical simulations of the considered model \cite{Timo,armour}. However, the
additional doublers ever appear in lattice propagator as it will be explained
below. Staggered fermion variables $\Psi$ are obtained via the spin
diagonalization: $ \psi_x = \Gamma_1^{x_1} \Gamma_2^{x_2} \Gamma_3^{x_3}
\Gamma_4^{x_4} \Psi_x$. Here always $x_3 = 0$. In terms of $\Psi$ the free
fermion action has the form:
\begin{equation}
S = \sum_x\Bigl( m\, \bar{\Psi}_x\Psi_x + \frac{1}{2}\sum_{i = 1,...,4}
[\bar{\Psi}_x \alpha_{xi} \Psi_{x + \hat{i}} - \bar{\Psi}_{x+\hat{i}}
\alpha_{xi} \Psi_{x }]\Bigr),\quad \alpha_{xi} = (-1)^{x_1 + ... + x_{i-1}}
\end{equation}
We keep the only component of $\Psi$. As a result the doublers play the role of
the components of the two original spinors. In order to reconstruct the
original spinor and flavor indices of the fermions we consider the lattice with
even number of lattice spacings in each direction. Let us subdivide the lattice
into the blocks consisted of elementary cubes. We denote the coordinates of the
blocks by $y_i$. Therefore, the coordinates of the lattice sites are $x_i = 2
y_i + \eta_i, \eta_i = 0,1$. We define the new fields \cite{Montvay}:
\begin{equation}
[\Phi_y]^{\alpha}_a = \frac{1}{8}\sum_{\eta} [\Gamma_1^{\eta_1}
\Gamma_2^{\eta_2} \Gamma_4^{\eta_4}]^{\alpha}_a \Psi_{2 y + \eta}
\end{equation}
Here index $\alpha = 1, ..., 4$ is the spinor index while $a = 1, ..., 4$ is
the flavor index. Matrices $\Phi$ have $4\times 4$ components. But not all of
these components are independent. We have the following constraint on $\Phi$: $
\Gamma_3 \Gamma_5 \Phi_y \Gamma_5 \Gamma_3  = \Phi_y $. There exists the
representation of gamma - matrices such that  the matrices $\Phi$ have the
form: $\Phi = \left(\begin{array}{cc}A&0\\
0 & B\end{array}\right)$. That's why we have two flavors of positive $ i
\Gamma_3 \Gamma_5 $ chirality and two flavors of negative $i \Gamma_3 \Gamma_5$
chirality that is two flavors of 4 - component Dirac spinors. Without
interactions in terms of $\Phi$ the propagator  in momentum representation (of
the blocked lattice) has the form \cite{Montvay}:
\begin{eqnarray}
 \tilde{\bf G}& = & -i \langle \bar{\Phi} \Phi \rangle = \Bigl( \sum_a \Gamma_a \frac{1}{2}{\rm sin}\, k_a - i (m -
\sum_a \frac{1}{2}(1 - {\rm cos}\, k_a)\Gamma_5 \otimes T_5 T_i)
\Bigr)^{-1}\nonumber\\ & = & \frac{ \frac{1}{2}\sum_a \Gamma_a {\rm sin}\, k_a
+ i (m - \frac{1}{2}\sum_a (1 - {\rm cos}\, k_a)\Gamma_5 \otimes T_5 T_i)
}{32[\sum_a \frac{1}{2}(1-{\rm cos}\, k_a) + m ^2]}
\end{eqnarray}
Here $T_i = \Gamma_i^T$ acts on the flavor indices while $\Gamma$ matrices act
on the Dirac indices. Momenta $k$ are $ k_1 = \frac{2\pi K_1}{N_x/2}\, \quad
k_2 = \frac{2\pi K_2}{N_y/2} \, \quad k_4 = \frac{2\pi K_4 + \pi}{N_t/2}, \quad
K_1,K_2,K_4 \in Z $. In this regularization the mass term is necessarily added.
At the end of the calculation one must set $m = 0$. This Green function turns
to the form (\ref{G4D}) with $\cal G$ in the form (\ref{Green0}) in the
continuum limit at $m = 0$. For $m=0$ the only pole of the Green function at $p
= 0$ appears. The fermion doublers do not have such poles. However, zeros of
the functions $g_a$, $a = 1,2,4$ appear at $p_a = \pi k_a, k_a \in Z$. At any
value of $m$ vector $n$ mentioned above has the following components: $ n_a =
g_a/\sqrt{g_a g_a},\, \quad g_1 =  i {\rm Tr}\,{\bf G}\, \Gamma_4 \Gamma_1/4\,
\quad g_2 =  i {\rm Tr}\,{\bf G}\, \Gamma_4 \Gamma_2/4\,\quad  g_3 =- i {\rm
Tr}\,{\bf G}/4$.
 Without interactions we have $n_a = \frac{{\rm sin}\, k_a}{\sqrt{\sum_a {\rm sin}^2
 k_a}}$.
From this expression we obtain $4$ monopole - antimonopole pairs in momentum
space placed in the positions of the fermion doublers.   For the surface that
encloses any of these points of the Brillouin zone we obtain the values ${\cal
N}_2 = \pm 1$. This demonstrates that the lattice formulation does not
eliminate  monopole in momentum space corresponding to the physical pole of the
Green function. However, this formulation also gives  monopoles that correspond
to the unphysical doublers.

When the interaction is switched on the practical prescription for the
calculation of the vector $n$ is $n_a(k)  =  g_a(k)/\sqrt{g_a(k) g_a(k)}, \quad
a = 1,2,4$ with
\begin{eqnarray}
g_a (k)& = & \frac{i}{16 N_1^2 N_2^2 N_t^2} \sum_{y,z} e^{i k (z -
y)}\sum_{\eta,\eta^{\prime} } (-1)^{\eta_1 +...+\eta_{a-1}}\nonumber
\\ &&\delta(\eta^{\prime}_i-[\eta_i + \delta_{ia}]{\rm mod} \, 2)\langle G(2 y + \eta, 2 z + \eta^{\prime})\rangle
\end{eqnarray}
Here $\langle G(2 y + \eta, 2 z + \eta^{\prime})\rangle$ is the staggered
fermion one - component propagator in the external field averaged over the
configurations of the $U(1)$ gauge field $A_4$ and over the pseudofermion
configurations (the latter give the fermion determinant in the averaging over
gauge fields).

Using expression (\ref{V}) we may calculate the value of $V\in SU(2)/U(1)$ at
any point of the momentum space lattice. Next, we may define the $U(1)$ gauge
field $B$ at any link of this momentum lattice via the following equation:
\begin{equation}
\left(\begin{array}{cc}{\rm cos} \phi  \,  e^{i B_{xy}} & {\rm sin} \phi \,
e^{i\chi}\\ -{\rm sin} \phi\, e^{-i\chi} & {\rm cos} \phi  \, e^{-i
B_{xy}}\end{array}\right) = V_x V^{\dag}_y
\end{equation}
The position of the monopole is given by $j = \frac{1}{2\pi}^* d [d B \, {\rm
mod} \, 2\pi]$. We expect that the pattern described above with $4$ monopole -
antimonopole pairs in momentum space will remain until a phase transition is
encountered.

\section{Conclusions and discussion}

In this paper we extend the construction of the topological invariant ${\cal
N}_2$  to  $\omega - {\bf p}$ space. The suggested construction works for the
case when the Coulomb interaction between the quasiparticles is present. We
also construct the gauge field in momentum space that has vanishing field
strength everywhere except for the poles and zeros of the Green function (and
the strings that connect them). The positions of poles and zeros themselves
coincide with the positions of monopoles extracted from the given gauge field.

The $8\times 8$ Green functions of the fermion quasiparticles are reduced to
$2\times 2$ matrices, and even further, to the elements of $su(2)$. These
matrices can be represented as $i{\cal G}\sigma_3 = g_3 \sigma_3 + {\bf g}_a
\sigma_a$ with real $g_3, {\bf g}_a$. The constructed topological invariant
catches zeros and poles of $\cal G$. If interactions  are absent, ${\cal N}_2 =
1$. When the Coulomb interaction is turned on, the equation ${\cal N}_2=1$
holds until the phase transition is encountered. This means that the pole in
$\cal G$ cannot disappear until the phase transition occurs.

The system may be transferred to various phases, where different symmetries of
the initial system are broken. There may appear different fermion condensates
\cite{Araki,rev_sym,Semenoff,Son,Timo}. The phase structure of the effective
field model of graphene is still unknown. Topology of momentum space must have
the relation to this phase structure. The constructed invariant ${\cal N}_2$
and the monopoles in momentum space have direct connection only to the phase
that includes the noninteracting system. The transition to the other phas(es)
may be accompanied with the change of ${\cal N}_2$. The transition between the
new phases may have connection to the other topological invariants. In
particular, the topological invariants for the $2+1$ gapped systems enter the
expression for the quantized Hall conductivity
\cite{VolovikBook,Hatsugai,HatsugaiHall,Hall_2008}. Also it is worth mentioning
that in the presence of the finite chemical potential the Fermi surface appears
that is related to the invariant ${\cal N}_1$ \cite{VolovikBook}.

The construction presented here is intended for the use mainly in lattice
simulation of the effective field theory of graphene at vanishing chemical
potential and in the absence of external fields (for the review of recent
numerical investigations of the model see \cite{Araki,Timo,armour} and
references therein). We expect that the phase transition(s) may take place to
the phase(s), where chiral symmetry of the noninteracting system is broken
\cite{rev_sym} in a certain way. The transition to the new phase must lead to
the change of the momentum space topology. The behavior of the monopoles in
momentum space constructed here are expected to be related intimately to the
mechanism of the phase transition(s) and to the nature of the new phase(s).
Their investigation may also be important for the understanding of the role of
doublers in various lattice discretizations of the Fermion systems.

The author kindly acknowledges private communication with G.E.Volovik, and
discussions with members of the lattice ITEP group M.I.Polikarpov,
P.Buividovich, V.I.Zakharov, O.Pavlovsky, M.Ulybyshev. This work was partly
supported by RFBR grants 09-02-00338, 11-02-01227, by Grant for Leading
Scientific Schools 679.2008.2. This work was also supported by the Federal
Special-Purpose Programme 'Cadres' of the Russian Ministry of Science and
Education, by Federal Special-Purpose Programme 07.514.12.4028.

\section{Appendix}

Let us denote $\bar{\chi} = \chi^{\dag} \sigma_3$. Then we introduce new
function $\tilde{\cal G}$ as  ${\cal G} = i \tilde{\cal G}\sigma_3$ and
represent it in the following form:
\begin{eqnarray}
i\tilde{\cal G}(x)  &= & \frac{1}{Z} \int D\bar{\chi} D\chi D A \bar{\chi}(0)
\chi(x){\rm exp} \Bigl( - \frac{1}{2}\int d^4x [\partial_{I} A_{4}]^2
\nonumber\\&& - \int d^3x \bar{\chi}([\partial_4 - i g A_4]\sigma_3 -
[\partial_1 ] \sigma_1 - [\partial_2 ] \sigma_2)\chi\Bigr)
\end{eqnarray}

We consider severar cases, when the transformational properties of the action
leads to symmetries of the Green function:

\begin{enumerate}

\item{} Let us consider the following transformation  $\chi \rightarrow i
\sigma_2[\bar{\chi}]^T$, $\bar{\chi} \rightarrow - i\chi^T \sigma_2,
A_4(x)\rightarrow A_4(-x)$ (remind that $\chi$ and $\bar{\chi}$ are independent
anticommuting variables), $x \rightarrow - x$. Using this transformation we
obtain:
\begin{eqnarray}
S_f  &= &   \int d^3x \bar{\chi}([\partial_4 - i g A_4]\sigma_3 - [\partial_1 ]
\sigma_1 -
[\partial_2 ] \sigma_2)\chi\nonumber\\
&\rightarrow & \int d^3x {\chi}^T\sigma_2([-\partial_4 - i g A_4]\sigma_3 +
[\partial_1 ] \sigma_1 +
[\partial_2 ] \sigma_2)\sigma_2\bar{\chi}^T\nonumber\\
&=& \int d^3x {\chi}^T([\partial_4 + i g A_4]\sigma_3 - [\partial_1 ] \sigma_1
+ [\partial_2 ] \sigma_2)\bar{\chi}^T\nonumber\\ &= &   \int d^3x
\bar{\chi}([\partial_4 - i g A_4]\sigma_3 - [\partial_1 ] \sigma_1 -
[\partial_2 ] \sigma_2)\chi
\end{eqnarray}

Measure over $\chi_{\pm}$ and the gauge field action are also invariant under
this transformation. As a result we obtain
\begin{eqnarray}
\tilde{\cal G}_{ab}(x) & = & \langle \bar{\chi}_a(0) \chi_b(x) \rangle =
\epsilon_{ac} \epsilon_{bd} \langle {\chi}_c(0) \bar{\chi}_d(-x) \rangle =
\epsilon_{ac}  \langle  \bar{\chi}_d(-x){\chi}_c(0)
\rangle \epsilon_{db}\nonumber\\
& = & \epsilon_{ac}  \langle  \bar{\chi}_d(0){\chi}_c(x) \rangle \epsilon_{db}
= - [\sigma_2 \tilde{\cal G}^T(x) \sigma_2]_{ab}
\end{eqnarray}

This implies $g_0(\omega, {\bf p}) = - g_0(\omega, {\bf p}) = 0$, $g_3(\omega,
{\bf p}) = g_3(\omega, {\bf p})$,  and ${\bf g}(\omega, {\bf p}) = {\bf
g}(\omega, {\bf p})$.

\item{} Analogue of CP - transformation corresponds to $\chi \rightarrow
\sigma_1[\bar{\chi}]^T$, $\bar{\chi}  \rightarrow  -\chi^T \sigma_1$, ${\bf x}
\rightarrow - {\bf x}, A_4(x_4,{\bar x}) \rightarrow -A_4(x_4,-{\bar x})$. In a
similar way we obtain:
\begin{eqnarray}
S_f  &= &   \int d^3x \bar{\chi}([\partial_4 - i g A_4]\sigma_3 - [\partial_1 ]
\sigma_1 -
[\partial_2 ] \sigma_2)\chi \rightarrow \nonumber\\
&\rightarrow & \int d^3x {\chi}^T\sigma_1([-\partial_4 - i g A_4]\sigma_3 -
[\partial_1 ] \sigma_1 -
[\partial_2 ] \sigma_2)\sigma_1\bar{\chi}^T\nonumber\\
&=& \int d^3x {\chi}^T([\partial_4 + i g A_4]\sigma_3 - [\partial_1 ] \sigma_1
+ [\partial_2 ] \sigma_2)\bar{\chi}^T\nonumber\\ &= &   \int d^3x
\bar{\chi}([\partial_4 - i g A_4]\sigma_3 - [\partial_1 ] \sigma_1 -
[\partial_2 ] \sigma_2)\chi
\end{eqnarray}
That's why
\begin{eqnarray}
\tilde{\cal G}_{ab}(x) & = & \langle \bar{\chi}_a(0) \chi_b(x) \rangle = -
\sigma^1_{ac} \sigma^1_{bd} \langle {\chi}_c(0) \bar{\chi}_d(x_4,-{\bar x})
\rangle = \sigma^1_{ac} \langle  \bar{\chi}_d(x_4,-{\bar x}){\chi}_c(0)
\rangle \sigma^1_{db}\nonumber\\
& = & \sigma^1_{ac}  \langle  \bar{\chi}_d(0){\chi}_c(-x_4, {\bar x}) \rangle
\sigma^1_{db} =  [\sigma_1 \tilde{\cal G}^T(-x_4, {\bar x}) \sigma_1]_{ab}
\end{eqnarray}
The Fourier transformation gives
\begin{equation}
\tilde{\cal G}(\omega, {\bf p}) =   \sigma_1 \tilde{\cal G}^{T}(-\omega, {\bf
p})\sigma_1
\end{equation}
Therefore, $g_0(\omega, {\bf p}) =  g_0(-\omega, {\bf p})$, $g_3(\omega, {\bf
p}) = -g_3(-\omega, {\bf p})$,  and ${\bf g}(\omega, {\bf p}) = {\bf
g}(-\omega, {\bf p})$.

\item{} Rotation of the ($1,2$) plane corresponds to the transformation $\chi
\rightarrow e^{i \phi \sigma_3/2} \chi$, and ${\bf x} \rightarrow e^{i \phi
\sigma_2} {\bf x}$ with the angle $\phi$. We have:
\begin{equation}
\tilde{\cal G}(\omega, {\bf p}) = e^{-i \phi \sigma_3/2} \tilde{\cal G}(\omega,
e^{i \phi \sigma_2}{\bf p})e^{i \phi \sigma_3/2}
\end{equation}
This implies $g_0(\omega, {\bf p}) = g_0(\omega, e^{i \phi \sigma_2}{\bf p})$,
$g_3(\omega, {\bf p}) = g_3(\omega, e^{i \phi \sigma_2}{\bf p})$,  and ${\bf
g}(\omega, {\bf p}) = e^{-i \phi \sigma_2}{\bf g}(\omega, e^{i \phi
\sigma_2}{\bf p})$.

\end{enumerate}

All mentioned above allow to derive the general form of $\cal G$:

\begin{equation}
g_0(\omega, {\bf p}) = 0, \quad g_3(\omega, {\bf p}) = \tilde{f}\Bigl(\omega,
|{\bf p}|^2\Bigr), \quad {\bf g}(\omega, {\bf p}) = {\bf p}
\tilde{h}\Bigl(\omega^2,|{\bf p}|^2 \Bigr)
\end{equation}
Here $\tilde{f}$ is odd as a function of $\omega$.

\end{document}